\documentclass[floatfix,11pt,showpacs,preprintnumbers,amsmath,amssymb]{revtex4}

\usepackage{epsf}
\usepackage{graphicx}  
\usepackage{dcolumn}   
\usepackage{bm}        

\newcommand{\be}{\begin{equation}}
\newcommand{\en}{\end{equation}}
 \newcommand{\bea}{\begin{eqnarray}}
 \newcommand{\ena}{\end{eqnarray}}

\begin{document}

\title{Does Unruh radiation accelerate the universe? A novel approach to the cosmic acceleration}
\author{Hongsheng Zhang\footnote{Electronic address: hongsheng@kasi.re.kr} }
\affiliation{
 Korea Astronomy and Space Science Institute,
  Daejeon 305-348, Korea }
 \affiliation{Department of Astronomy, Beijing Normal University,
Beijing 100875, China}
\author{Hyerim Noh }
 \affiliation{
 Korea Astronomy and Space Science Institute,
  Daejeon 305-348, Korea }
 \author{Zong-Hong Zhu}
  \affiliation{
 Department of Astronomy, Beijing Normal University, Beijing 100875, China}
 \author{Hongwei Yu}
 \affiliation{Department of Physics and Institute of  Physics,
 Hunan Normal University, Changsha, Hunan 410081, China}

\date{ \today}

\begin{abstract}
 We present a novel mechanism for the present acceleration of the universe.
  We find that the temperature of the Unruh radiation perceived by
  the brane is not equal to the inherent temperature (Hawking temperature
 at the apparent horizon) of the brane universe in the frame of Dvali-Gabadadze-Porrati (DGP) braneworld
 model.
   The Unruh radiation perceived  by a dust
dominated brane is always warmer than the brane measured by the
geometric temperature, which naturally  induces an energy flow
between bulk and brane based on the most sound thermodynamics
principles.
   Through a thorough investigation to the microscopic mechanism of interaction between bulk Unruh radiation and
   brane matter, we put forward that an energy influx from bulk Unruh radiation to the dust matter on the brane
     accelerates the universe.

\end{abstract}


\preprint{arXiv: 0908.1001}
 \maketitle

\newpage

\section{Introduction}
    An unexpected discovery appeared in 1998, that is, our universe is accelerating rather than decelerating \cite{acce}.
     This discovery has significant and far-reaching influence on both fundamental physics and astronomy.
     The invisible sector of the universe (or the modified terms to Einstein gravity) dominates the present
      evolution and determines the final destiny of the universe. To understand its physical nature,
      we have to go beyond the standard model.
      Various explanations of this acceleration have been proposed; see \cite{review}  for recent reviews with fairly
       complete lists of references of different models. However, the physical nature of the cosmic acceleration remains as a
       most profound problem in sciences.

      In different approaches to the cosmic acceleration problem,
      braneworld models draw much attention  in recent years.
      Inspired by string/M theory the braneworld gravity
   is introduced for several problems in standard model, firstly for the hierarchy
   problem. The braneworld gravity has been extensively studied, for a review, see\cite{rev}.
    In the braneworld scenario
   the standard model particles are confined to a brane, while
   gravity can propagate in the whole space time.  At the
   low energy region general relativity recovers, while at the high energy
   region the behavior of gravity is strongly modified.  This may yield
   remarkable changes in gravity dynamics with several different implications for cosmology,
   black holes and high energy physics.
     Among various braneworld models, the model
 proposed by Dvali,
 Gabadadze and Porrati (DGP)~\cite{dgpmodel} is impressive and a leading one in cosmology.
 In the DGP model, the bulk is a flat Minkowski spacetime, but a reduced
 gravity term appears on a tensionless brane.
 In this model, gravity appears
 4-dimensional at short distances but is altered at distance large
 compared to some freely adjustable crossover scale $r_0$ through the
 slow evaporation of the graviton off our 4-dimensional brane world
 universe into an unseen, yet large, 5th dimension.  At short distances the 4-dimensional
 curvature scalar $R$ dominates and ensures that gravity looks
 4-dimensional. At large distances the 5-dimensional curvature scalar
 ${}^{(5)}R$ takes over and gravity spreads into extra dimension. The late-time
 acceleration is driven by the manifestation of the excruciatingly
 slow leakage of gravity off our four-dimensional world into an extra
 dimension. This offers an alternative explanation for the current
 acceleration of the universe \cite{dgpcosmology}. However, just as LCDM model, it is
 also suffered from fine-tune problem. In LCDM model, $\Lambda$ is
 an unimaginatively tiny constant compared to the expectation of the vacuum if our quantum field
 theory (QFT) is valid till to Planck scale. We know QFT is at least
 valid at eletro-weak scale, whose expectation of zero-point energy is still $10^{60}$
 higher than the scale of the cosmological constant.
  To solve the dark energy problem
 in frame of
  DGP  model, people require
  $r_c\sim H_0^{-1}$, where $r_c$ is a constant, and $H_0$ is the present Hubble parameter.
   Similarly, we can
  ask why $r_c$ is so large that it is  approximately equals the Hubble
  radius.

  Interaction is a universal phenomenon in the physics world. An
  interaction term between bulk and brane has been invoked as a possible mechanism for
  the cosmic acceleration in frame
  of braneworld model \cite{interbrane}. However, in previous works little attention has
  been devoted to the physical mechanism of such a term. There is a natural physical origin for this
  interaction which has not been adverted before, that is, Unruh effect.
   In the braneworld scenario a brane is moving in the bulk, and
   generally speaking, its proper acceleration does not vanish. Thus,
    a brane, qua an observer in the bulk space, should
   perceive the Unruh radiation. The Unruh effect of RS II braneworld
   is investigated in \cite{jenn}.

    Even for a Minkowski bulk, the
   brane will sense a thermal bath filled with radiations. If the
   temperature is different from the inherent temperature of the
   brane, energy flow will appear based on the most sound principles
   of thermodynamics. There are several different particles are
   confined to the brane, which are not in thermal equilibrium.
   Then, what is the inherent temperature of the brane in the frame
   of gravity theory? It will be useful to give a brief account of
   the previous approaches, especially on black hole thermodynamics.

    The relation between gravity theory and thermodynamics is an interesting and
   profound issue. The key quantities bridging the gravity and thermodynamics are
   temperature and entropy.  Temperature of an ordinary system denotes the the average kinetic energy
   of microscopic motions of a single particle. To gravity the temperature becomes subtle. Since we do not
   have a complete quantum theory of gravity, for general case we can not use the
   usual way to get the temperature of the gravity field. Under this situation people
   set up some thermodynamics and statistical quantities of gravity by
   using semi-quantum (matter field is quantized, but gravity remains classical)
   theory, though the concept of gravitational particle is not clear.  The black hole
   thermodynamics (in fact, spacetime thermodynamics, because  the physical quantities in black hole thermodynamics
    should be treated as the quantities of the globally
   asymptomatic flat manifold) is set up in \cite{4law} and confirmed by Hawking radiation \cite{hawk}.

   In the initial work \cite{4law}, the temperature of the spacetime is suspected by an analogy to the ordinary
   thermodynamics. Shortly, it is recognized that the temperature is just the temperature of the
   radiation emitted from the black hole by a semi-quantum treatment.
   The method in \cite{hawk} depends on the global structure of the spacetime. The vacuum state is treated as nonvacuum state by
   observers at spacelike infinity, which roots in the fact that vacuum state around the black hole is different from the vacuum at
   spacelike infinity. Along this clue Unruh found that a uniformly accelerating particle detector with proper
  acceleration $A$ in Minkowski vacuum
  perceives different excited modes exactly in Bose-Einstein
  distribution with a temperature $T=A/2\pi$ \cite{unruh}.
  Unruh effect helps us to derive the local
   temperature of a gravitational system.  Unruh effect has
  inherent relations to Hawking effect.  After
   a conformal transformation, the thermal particles detected by an
   accelerated detector becomes thermal particles seen by an
   inertial observer in curved spacetime, which is just the Hawking
   effect in the case of black hole \cite{bril}.

   The laws of black hole mechanics are results of classical
   Einstein gravity, for qualification of laws of thermal dynamics
   the quantum theory is required. The mathematical form keeps
   invariant. This implies that the classical gravity theory may
   hide information from quantum theory. This possibility is
   investigated in \cite{jaco}. The Einstein equation is  reproduced from the
   proportionality of entropy and horizon area together with
   the first law of thermal dynamics, $\delta Q = T
   dS$,  jointing to heat, entropy, and temperature, where the temperature is the Unruh
   temperature to an observer just behind a causal Rindler horizon.
   The entropy is supposed to be proportional to the area of this
   horizon and the heat flow is measured by this observer. The
   significance of this derivation is that the Unruh effect is a
   result of quantum field theory but the derivation of classical
   Einstein theory depends on it.

   More directly, for a dynamical
   spacetime, a similar procedure reproduces the Friedmann equation.
   In this case
   one should apply the first law of thermodynamics to the trapped surface (apparent horizon) of an FRW
  universe and assume the geometric entropy given
  by a quarter of the apparent horizon area and the temperature given by the inverse
  of the apparent horizon \cite{cai}. There are several arguments that the apparent horizon
  should    be a causal horizon and is associated with the gravitational
  entropy and Hawking temperature \cite{bak}. So, for an expanding
  universe, the Hawking temperature at the apparent horizon should
  be treated as the inherent temperature of the universe. For a
  brane universe its inherent temperature may be higher or lower than
  the temperature of the Unruh radiation in the bulk, which thus
  triggers an energy flow between brane and bulk. We will study this
  possibility in frame of DGP braneworld model.

  This paper is organized as follows: In the next section we give the
  basic construction of DGP model and present the Unruh temperature
  perceived by an inertial observer (inertial respective to the brane, which we will explain in detail)
  on the brane. In section III, we
  investigate the condition for the temperature to be valid. In section IV, we study
   geometric temperature of a DGP brane in detail and
  find the thermal equilibrium condition for the Unruh radiation and
  the brane (evaluated by its geometric temperature). In section V, we present a detailed
  study of the energy exchange between bulk Unruh radiation and the brane matter by statistical mechanics.
   Our conclusion
  and discussion appear in section VI.

\section{Unruh effect for a DGP brane}

  In this section we discuss the Unruh effect for a DGP brane.
  Before studying the Unruh effect of a brane, we shall first give a
   brief review of the gravitational thermodynamics and Unruh
   effect in 4 dimensional theory, especially the reality of this significant effect.

  Unruh effect states that the concept of particle depends on observer.  This
  amazing effect generates some puzzles, sometimes even treated as paradoxes.

  First, Unruh's original construction is not consistent
  because his quantization is not unitarily equivalent to the standard
  construction associated with Minkowski vacuum. Hence some authors used
   mathematically more rigorous methods to solve this problem \cite{math}.

   Second, the temperature of an Unruh particle detector, which is in thermal
    equilibrium with Unruh radiation it experiences in Minkowski vacuum,
    is higher than the temperature
    of the vacuum to an inertial observer. Hence, does it $really$ emit  radiation
    for an inertial observer, just like an accelerated charged particle?
   It was argued that there was no radiation flux from an Unruh observer
   \cite{no}. Unruh made an almost the same calculation as in \cite{no}, but he
   found  extra terms in the two point correlation  function of the field
   which would contribute to the excitation of a detector \cite{1992}. It was
   pointed out that the extra terms were missing in \cite{no}, which shape a
   polarization cloud about an Unruh detector \cite{cloud}. The above analysis
   and detailed analysis (including non-uniformly accelerated observers) in \cite{non}
   showed that in a 2-dimensional toy model, there is no radiation flux from
   the detector. Recently it was found  that there exists a positive
   radiated power of quantum nature  emitted by the detector in 4 dimensional
   space \cite{lin}.

   It should be noted that the  response of an Unruh observer to the Minkowski vacuum is
   independent of its inner structure: The distribution of the
   different excited modes perceived by the observer depends only on
   the acceleration of the observer.
   This implies that the Unruh temperature is an
   inherent property of the quantum field. The detector only plays
   the role of test particles. This effect still exists even without
   a detector, which can be seen clearly from the derivation by Bogolubov
   transformation. The Unruh effect can be also derived in terms of
   the spontaneous excitation of accelerating atoms~\cite{Audretsch94,ZYL06,ZY07}.

   Several contemporary research themes, for example, the black hole thermodynamics \cite{wald}, the quantum entanglement state
   \cite{yuhan}, the Lorentz symmetry breaking \cite{kkrs}. For
   extensive references  of the Unruh effect, see a recent review article
   \cite{ureview}.

  Before studying the Unruh effect for a DGP brane we warm up by a short review the DGP brane.
   The basic construction of DGP model can be written as follows \cite{dgpmodel},

 \be
 \label{action}
 S=S_{\rm bulk}+S_{\rm brane},
 \en
where
 \be
 \label{actbu}
  S_{\rm bulk} =\int d^5x \sqrt{-{}^{(5)}g}
 \left[ {1 \over 2 \kappa^2} {}^{(5)}R  \right],
 \en
and
 \be
 \label{actbr}
 S_{\rm brane}=\int d^4 x\sqrt{-g} \left[
{1\over\kappa^2} K^\pm + L_{\rm brane}(g_{\alpha\beta},\psi)
\right].
 \en
 Here $\kappa^2$ is the  5-dimensional gravitational constant,
 ${}^{(5)}R$ is the 5-dimensional
 scalar curvature. The induced metric $g_{\mu\nu}$ is defined as
 \be
 g_{\mu\nu}={}^{(5)}g_{\mu\nu}+n_{\mu}n_{\nu},
 \en
 where the lower case of
 Greeks run from $0\sim 3$, and $n_\mu$ is the normal to the brane.
  $x^\mu ~(\mu=0,1,2,3)$ are the induced 4-dimensional
 coordinates on the brane, $K^\pm$ is the trace of extrinsic
 curvature on either side of the brane and $L_{\rm
 brane}(g_{\alpha\beta},\psi)$ is the effective 4-dimensional
 Lagrangian, which is given by a generic functional of the brane
 metric $g_{\alpha\beta}$ and matter fields $\psi$ on the brane. In
 this article we adopt a mostly negative signature.
 For DGP model, an induced 4 dimensional Ricci scalar term appears in the brane
 Lagrangian $L_{\rm brane}$,
\bea
 \label{lagbr}
 L_{\rm brane}=  {\mu^2 \over 2} R +  L_{\rm
 m},
 \ena
  where $\mu$ is the reduced 4 -dimensional Planck mass, $R$
 denotes the  scalar curvature on the brane,
  and $L_{\rm m}$ stands for the Lagrangian of
 matters on the brane.

 Assuming a Friedmann-Robertson-Walker (FRW) metric on the brane, we
 can
 derive the Friedmann equation on the brane \cite{dgpcosmology} (for a more extensive study of the DGP cosmology in which a
 Gauss-Bonnet term a Weyl term appear in the bulk, see
 \cite{gbzhang} ),
 \be
 \label{fried}
 H^2+\frac{k}{a^2}=\frac{\rho}{3\mu^2}+\frac{2}{r_c^2}+\frac{2\epsilon}{r_c}
 \left(\frac{\rho}{3\mu^2}+\frac{1}{r_c^2}\right)^{1/2},
 \en
 where $\rho$ denotes matter energy density on the brane,
 $r_c=\kappa^2\mu^2$, denotes the cross radius of DGP brane,
 $\epsilon=\pm 1$, represents the two branches of DGP model, $H$ is Hubble parameter, $k$ is spatial curvature, and $a$
 is the scale factor of the brane.
   The branch $\epsilon=1$ was treated as an unstable branch based on the linear order perturbations.
  However, a recent study shows that ghosts do not appear under small fluctuations of an empty
  background, and conformal sources do not yield instabilities either \cite{stable}.

 Now we begin to investigate the Unruh effect for a DGP brane in a
 Minkowski bulk. We work in a bulk-based coordinate system,
 \be
 dS^2=d\hat{T}^2-dR^2-R^2d\Omega_3^2,
 \label{5dimmetric}
 \en
 where $dS^2$ is the 5 dimensional line-element, $d\Omega_3^2$
 denotes a 3 dimensional sphere.
   A free falling observer on the brane, which is just the comoving observer in the FRW universe,
  can be described in the FRW coordinates,
  \be
  x^{\mu}=(t,x^1_0,x^2_0,x^3_0),
  \en
  where $t=x_0$ is the standard world time function, and
  $x^1_0,x^2_0,x^3_0$ are three coordinates of the unit 3-sphere
  $\Omega_3$. For a comoving observer, they are constants. Hence, clearly its acceleration
  is zero in the brane-viewpoint. However, it accelerates in
  the cross direction of the brane, or the direction of
  extradimension. The observer moves along a non-geodesic orbit
  measured by the full 5-dimensional metric. This is the exact
  meaning of ``A brane accelerates in the bulk", and the
  acceleration of the brane indicates the acceleration of such
  comoving observers.

  We assume a free-falling particle on the brane, as a particle detector,  moves along the worldline
  \be
  x^{B}(t)= (\hat{T},R,x^1_0,x^2_0,x^3_0),
  \label{orbit1}
  \en
 where the capital Latin letters run from $0\sim 4$.
 For a moving brane in the $R$ direction, the velocity of a free falling observer on the brane reads
 \be
 U^{A}=(\dot{\hat{T}}, \dot{R},0,0,0),
 \en
 which satisfies the normalization condition,
 \be
 U_{A}U^{A}=1.
 \en
  The induced metric on the brane reads,
  \be
  {ds}^2=dS^2-U\otimes U=dt^2-a(t)^2d\Omega_3^2,
  \en
  where we have labeled the radius coordinate $R$ by a new symbol
  $a$, which is more frequently used in cosmology.

 The unit normal of the brane is vertical to the
 velocity and spacelike,
 \be
 n_{A}U^{A}=0,~n_{A}n^{A}=-1.
 \en
 The acceleration of the brane is defined as,
 \be
 A^{C}=U^{B}\nabla_{B} U^{C}.
 \en
 We consider the component of the extrinsic curvature along the
 velocity,
 \be
 K_{tt}=K_{AB}U^AU^B=-n_CA^C.
 \en

   By using the junction condition
 across the brane
 \be
 [K_{\mu\nu}] = - \kappa ^2 ( s_{\mu\nu} -\frac{1}{3} g_{\mu\nu} s),
 \en
 where $[K_{\mu\nu}]$ denotes the difference of the extrinsic
 curvatures of the two sides of the brane, $s_{\mu\nu}$ represents the
 energy-momentum confined to the brane, we derive the amplitude of the acceleration
 of  brane \cite{lang},
 \be
 A=\left|^{(5)}g_{CD}A^CA^D\right|^{1/2}=\frac{\kappa ^ 2}{6}\left|2\rho_e + 3p_e\right|.
 \label{acce}
 \en
 Here $\rho_e$ and $p_e$ are the effective energy density and pressure of
 the brane, which are defined as
 \bea
 \label{rhoe}
  \rho_e=s_0^0,\\
  p_e=-s_i^i,
  \label{pe}
   \ena
  where $i$ is an arbitrary spatial index and we need not sum over the
  index $i$. And $s$ is the effective energy momentum confined to
  the brane,
  \be
  s_{\mu\nu}\triangleq \frac{2}{\sqrt{-g}}\frac{\delta({\sqrt{-g} L_{\rm
  brane}})}{\delta g_{\mu\nu}},
  \label{embrane}
  \en
  where $L_{\rm
  brane}$ is given by (\ref{lagbr}). Due to the above equation, the
  energy momentum $s_{\mu\nu}$ includes the contribution of the
  matter term $L_m$, as well as the induced curvature term $R$ on the brane. That is the reason why we
  call the density and pressure in (\ref{rhoe}) and (\ref{pe})
  effective density and pressure. We stress that the effective
  density $\rho_e$ is different from the local density $\rho$ of the
  matter confined to the brane in (\ref{fried}), which does not
  include the geometric effect of the induced Ricci term $R$.

  We see that, generally
 speaking,
  $A$ does not vanish. So the brane may perceive Unruh-type
  radiation in the bulk.
To discuss Unruh radiations seen by the
 observer on the brane, we introduce a 5-dimensional scalar field in the bulk,
 which is in its Minkowski vacuum state,
 \be
 {\cal L}_{\phi}=\frac{1}{2}\partial_{A}\phi\partial^A\phi-\frac{1}{2}m^2\phi^2,
 \en
 where  $m$ denotes $\phi$'s mass parameter.
 Since the scalar is in its vacuum state, classically it does no work
 on the brane and bulk dynamics.

  Several different methods have been proposed
  to derive Unruh effect since Unruh's original work \cite{bril}.
  Here we use the Green function method.
  We consider a detector minimally coupled to a 5-dimensional scalar field
  $\phi$ in the bulk
  \footnote{One often adds a conformal coupling term to gravitational field  in
  the action $\L_{{\rm con}}=-\frac{1}{2}\xi R\phi^2$. But there are some subtles in this couple, because there
  exists two Ricci scalars in the DGP model and two types of
  conformal transformations. Wether the Lagrangian is conformally
  invariant depends on definition. Here we just consider the minimal
  coupling case.}.

  The  Lagrangian $cm(t)\phi[x(t)]$ describes the interaction
  between detector and field, where $c$ is the coupling constant and
  $m$ denotes the moment operator of the detector.
  It is shown in \cite{bril} that, for a small number $c$,
   the probability amplitude of the  transition
  from the ground state $|E_0>$ of a particle detector coupled to a scalar
  field in its vacuum state $|0_M>$ to an excited state $|E>$, where $M$ stands for Minkowski,
  reads
 \be
 c^2 \sum_E |<E|m|0_M,E_0>|^2  {\cal F}(E-E_0),
 \en
  where the detector
 response function ${\cal F }(E)$  is defined as
 \be
 \label{respo}
 {\cal F}(E-E_0) = \int^\infty_{-\infty} dt \int^\infty_{-\infty}
 dt'G^+(x(t),x(t'))e^{-i(E-E_0)(t-t')}
 ,
 \en
 and the Wightman  Green function $
 G^+(x,x')$ is defined as
 \be
  G^+= <0_M|\phi (x) \phi (x') |0_M>.
  \label{define}
  \en
 We note that the term $ |<E|m|0_M,E_0>|^2$ depends on the structure
 of the particle detector, but ${\cal F }(E-E_0)$ does not, which
 reflects the inherent properties of quantum fields.
 It is clear that for a general trajectory the response function is not zero.

 For a uniformly accelerated trajectory the Wightman Green function becomes a function of
  $\Delta t =t - t '$. Since the detector will detect infinite particles
  in its whole history,  the response
 function (\ref{respo}) is not well-defined. Under this situation it will make sense to consider
 the unit response function, which describes in unit time interval,
 \be
  {\cal U}(E-E_0)= \int^\infty_{-\infty} d \Delta t e^{-i(E-E_0)\Delta t }  G^+(\Delta t ) .
 \label{unit}
 \en
  Since $x_0^1,x_0^2,x_0^3$ are constant, we just set
  $x_0^1=x_0^2=x_0^3=0$ after a coordinates transformation. For a
  uniformly accelerated observer in the cross-brane direction,
  the coordinates in (\ref{orbit1}) read,
  \be
  x_0^1=x_0^2=x_0^3=0,~\hat{T}=\frac{\sinh (tA)}{A},R=\frac{\cosh
  (tA)}{A},
  \label{orbit}
  \en
  where $A$ is a constant, denoting the magnitude of acceleration of the
  detector.
  Then, integrating (\ref{define}) directly with proper boundary condition (corresponding to
   proper contour \cite{bril}), we obtain its concrete form in
   McDonald function (Bessel function with imaginary arguments),
   \be
   G^+(\Delta t)=\frac{e^{i5\pi/4} (mA)^{3/2}}{16\pi^{5/2}(\sinh(A\Delta t/2-i\varepsilon))^{3/2}}
   K_{3/2}\left(\frac{i2m}{A}\sinh(A\Delta t/2-i\varepsilon)\right),
 \en
 where, as usual, $\varepsilon$ is a small positive number. The Wightman Green function for massless mode reads,
 \be
 D^+({\Delta t})=\lim_{m\to 0} G^+(\Delta
 t)=-\frac{i}{64\pi^2}\frac{A^3}{(\sinh (A\Delta
 t/2-i\varepsilon))^3}.
 \en
 Generally speaking, Unruh effect is a very weak effect. For example, in 4 dimensional
 Minkowski space, it needs about an acceleration $10^{21}$ meter per second$^2$ to increase temperature 1 K, hence various
 massive modes are difficult to be excited.  The
 mode with zero mass is the easiest mode to be excited. Therefore, we consider  excitations of massless modes.
  Substituting $D^+({\Delta t})$ into (\ref{unit}), one obtains the unit response
 function  with a contour closed at lower half plane of complex $\Delta t$ by using the Jordan's lemma,
 \be
 {\cal U}(E-E_0)=\frac{1}{32\sqrt{\pi}}
 \frac{A^2+4(E-E_0)^2}{e^{2\pi (E-E_0)/A}+1}.
 \label{resp}
 \en
 This is the response function for a particle detector confined to the FRW brane, and comoving to the FRW
 brane, which describes photon gas system at temperature $A/2\pi$.
 We see that a Fermi-Dirac factor appears, which implies that the particles of Rindler radiation behaves
 as Fermions, though all the
 excited modes we integrated to derive the response function are Bosonic. It is not a phenomenon
 completely new. In 1986, Unruh pointed out that the Fermi-Dirac
 factor would appear in the response function for an accelerated
 monopole of a massless field in an odd number of space dimension,
 arising from integration over all modes for a scalar field  \cite{unruh}.

  In the above text of this section, only a brane with  positive
  spatial curvature is discussed. A spatially flat brane can be
  treated as a limit when $R$ is large enough. To the case of a
  negative spatial curvature, we need to replace the
  3-sphere by a 3-hyperbola $H_3$ in (\ref{5dimmetric}). A
  5-dimensional Minkowski space sliced with space-like 3-hyperbola can be written as,
 \be
 dS^2=-d\hat{T}^2+dR^2-R^2dH_3^2.
 \label{5dimmetric1}
 \en
   Then the following discussions exactly follow the case of positive spatial
  curvature. The resulting distribution function (\ref{resp}) is still valid.

 \section{Quasi-stationary acceleration stage}

 Up to now our analysis is limited to uniformly accelerated
 detectors. But from (\ref{acce}) we see that the acceleration of
 the  brane is not a constant. Physically, the formula ($\ref{resp}$) remains
 valid when the acceleration varies slowly. Mathematically we can estimate
 the time over which the constant
 acceleration approximation is valid by imposing that the variations in the acceleration are
  small, e.g., expanding the acceleration around some time $t_0$ to the first order
  $t-t_0$:
  \be
  A(t)=A(t_0)+(t-t_0)\frac{dA}{dt}\left|_{t=t_0}\right.
  \en
  Here, we adopt the cosmic time (proper time) $t$ rather than the proper time
  coordinate time $\hat{T}$ of the detector since the cosmic time is directly related
  to our measurement of cosmological parameters.
  For evaluating the variation of the acceleration, define characteristic time
  \be
  t_c\triangleq \left|\frac{A(t_0)}{dA/dt |_{t=t_0}}\right|,
  \label{tc}
  \en
  which is a function of $t_0$, i.e., it varies with the evolution
  of the universe. Our condition for ``acceleration varies slowly''
  requires that the time scale of a physical process we concerned  is much less than the characteristic time,
  \be
  t_i\ll t_c.
  \en
  The time scale of a physical process $t_i$ must be shorter than the age of the universe.
  So if the characteristic time
  $t_c$ is much larger than the time scale of the universe, we can
  safely treat the acceleration as a constant for any physical
  processes.

  Here we present some examples to explain this condition. We work
  in a frame of ``$r_c$CDM" model, that is, a DGP brane universe
  filled with dust, whose density function $\rho$ in Friedmann
  equation (\ref{fried}) only includes pressureless matter, with about $30\%$ of the critical density.

  First, we consider the inflationary phase at the early universe.
  At such a high energy scale, the infrared correction of DGP model
  to the standard general relativity,
  which becomes important at late time universe, can be omitted
  safely. So the effective energy density and pressure in
  (\ref{acce}) are just the ordinary density and pressure of the
  universe. At the whole inflationary phase, the density of the universe
  is approximately constant, and the pressure $p=-\rho$. Hence,
  \be
  \frac{dA}{dt}=-\frac{\kappa^2}{6}\frac{d\rho}{dt}\sim 0,
  \en
  that is, from (\ref{tc}), the characteristic  time is very long.
  In this case, the Unruh temperature is well defined.

  Second, the universe enters a radiation dominated phase after the
  inflation. At this stage, the DGP-correction to the standard
  model is still tiny. Thus, we adopt the the same approximation as
  above, $\rho_e=\rho,~p_e=p$. For a radiation dominated universe,
  we have
  \be
  p=\frac{1}{3}\rho.
  \label{rhop}
  \en
  The continuity equation reads,
  \be
  \frac{d\rho}{dt}+3H(\rho+p)=0.
  \label{conti1}
  \en
 Substitute (\ref{rhop}) and (\ref{conti1}) into (\ref{tc}), we
 reach,
  \be
  t_c=\frac{\sqrt{3}}{4}\frac{\mu}{\sqrt{\rho}},
  \en
  from which we see that the characteristic time becomes longer when
  the energy density becomes lower.
 For instance, when $\rho=(0.1$Mev)$^4$, we get $t_c\sim 1$s,
 which is much shorter than the age of the universe at that time.

 And then the universe is diluted to be thinner and thinner. The DGP correction becomes
 more and more important. Under this situation we have to introduce
 the correction terms. By using (\ref{embrane}) (see also
 \cite{eff}), we derive
 \bea
 \label{effrho}
 \rho_e=\rho-\mu^2(3X),\\
 p_e=p+\mu^2(\frac{2\ddot{a}}{a}+X),
 \label{effp}
 \ena
 where $X$ is defined as
 \be
 X\triangleq H^2+k/a^2.
 \label{xdefi}
 \en
 Substituting  to the Friedmann equation (\ref{fried}), we obtain
   \be
 X= \frac{\rho}{3\mu^2}+\frac{2}{r_c^2}+\frac{2\epsilon}{r_c}
 \left(\frac{\rho}{3\mu^2}+\frac{1}{r_c^2}\right)^{1/2}.
 \label{xrho}
 \en

 As for standard cosmology, $\frac{2\ddot{a}}{a}$ can be obtained
 from the Friedmann equation  and the continuity equation
 . Here, similarly, from Friedmann equation
 (\ref{fried}) and the continuity equation (\ref{conti}), we derive
 \be
 \frac{\ddot{a}}{a}=X-\frac{1}{2\mu^2}\frac{\rho+p}{1+\frac{\epsilon
 }{\sqrt{X}}\frac{1}{r_c}}.
 \label{aa}
 \en
 Then substituting (\ref{effrho}) , (\ref{effp}) and (\ref{acce}) into (\ref{tc}) ,
 we arrive at the following rather complicated from,
  \be
 t_c=\frac{1}{3}\left[2+\frac{3\mu^2X}{\rho}-{3}{(1+\frac{\epsilon}{r_c\sqrt{X}})^{-1}}\right]
 (X-\frac{k}{a^2})^{-1/2}V^{-1},
 \en
 where
 \be
 V=3-\frac{1}{3}(1+\frac{\epsilon}{r_c\sqrt{X}})^{-1}+\frac{\epsilon}{r_c}\left(\frac{\rho}{3\mu^2}+\frac{1}{
 r_c^2}\right)^{-1/2}-\frac{1}{2}\frac{\rho}{r_c^2\mu^2
 X^{3/2}}\left[1+
 \frac{\epsilon}{r_c} \left(\frac{\rho}{3\mu^2}+\frac{1}{
 r_c^2}\right)^{-1/2}\right].
 \en
  Substituting the current parameter of the universe $\rho=(1.8\times
  10^{-3}$ev$)^4$ and we suppose that the universe is spatially
  flat, then we obtain, either for the negative or the positive branch,
   $t_c\sim 10^{17}$s, which is at the same scale of the age of the universe.
  So the constant acceleration is a perfect approximation under
  this condition.

 Then, at least of late time universe, we can safely use (\ref{resp}), which is a thermal distribution
 at a temperature of
  \be
 T=\frac{A}{2\pi}.
 \label{tem}
 \en

 Associating with (\ref{acce}), we get
 \be
 T=\frac{\kappa^2}{12\pi}|2\rho_e+3p_e|.
 \label{temp}
 \en

 \section{compared to geometric temperature of the brane}
 We see that the temperature of Unruh radiation the brane perceives,
 as
 displayed in (\ref{temp}),  only relates to the energy density and
 pressure of the brane for a given DGP model. Then, a natural question emerges: Is the Unruh radiation
 hotter, colder or at the same temperature to the brane? If there is energy exchange
 between bulk and brane, we also need the temperature of the brane
 to  decide the direction of the energy flux. But there are several
 different particles on the brane. They had gone out of thermal equilibrium long
 before. Hence we'd better to find the characteristic temperature of
 the braneworld independent of its detailed microscopic
 construction. As we have pointed out Unruh radiation is a geometric feature of
 the space, which is unrelated to the construction of the detector.
 Just as well we need the characteristic temperature of the brane.

 We know that the formulae of black hole entropy
 and temperature have a certain universality in the sense that the
 horizon area and surface gravity are purely geometric quantities
 determined by the space geometry, once Einstein equation
 determines the space geometry. As we have mentioned above, Einstein equation can be reproduced
 by thermal dynamics considerations.  As for the case of cosmology,
 applying the first law of thermodynamics to the apparent horizon of an FRW
  universe and assuming the geometric entropy given
 by a quarter of the apparent horizon area and the temperature given by the inverse
 of the apparent horizon, the Friedmann equation can be derived.
 This celebrated result implies that the inverse number of the
 apparent horizon is the geometric temperature of the universe,
 which is independent of the microscopic structures of the particles confined to the
 brane.  The apparent horizon in the dynamical universe is
  a marginally trapped surface with vanishing
 expansion. Straightforward calculation yields the radius of the apparent horizon
 \be
 R_A=X^{-1/2},
 \en
 where $X$ is defined in (\ref{xdefi}).
       And then the geometric temperature of
 the brane reads
 \be
 T'=\frac{R_A^{-1}}{2\pi}=\frac{X^{1/2}}{2\pi}=\frac{1}{2\pi}\sqrt{
 \frac{\rho}{3\mu^2}+\frac{2}{r_c^2}+\frac{2\epsilon}{r_c}
 \left(\frac{\rho}{3\mu^2}+\frac{1}{r_c^2}\right)^{1/2}}~~.
 \label{gt}
 \en

  Substituting (\ref{aa}) into (\ref{effp}) and (\ref{effp}) and (\ref{effrho}) into (\ref{temp}), we
 arrive at
  \be
  T=\left|\frac{r_c}{12\pi\mu^2}\left(2\rho+3p\right)+\frac{r_c}{4\pi}\left(X-\frac{1}{\mu^2}
  \frac{\rho+p}{1+\frac{\epsilon
 }{\sqrt{X}}\frac{1}{r_c}} \right)\right|,
 \label{ut}
  \en
 where $X$ is given by (\ref{xrho}). Therefore we obtain the
 explicit form
 temperature of the brane, which is determined by its energy density and
 pressure. By contrast the geometric temperature only depends on the energy density, which
 has no relation to the pressure. We see that both Unruh temperature (\ref{ut}) and geometric
 temperature (\ref{gt}) are unrelated to the spatial curvature of the brane.

 To  investigate further the evolution of the Unruh temperature and
 the geometric temperature during the history of the universe, we
 write them in dimensionless form,
 \be
 \frac{T'}{H_0}=\frac{X^{1/2}}{2\pi H_0}=\frac{1}{2\pi}\left[ x\Omega_m+2\Omega_{r_c}+2\epsilon
 \sqrt{\Omega_{r_c}} (x\Omega_m+\Omega_{r_c})^{1/2}\right] ^{1/2},
 \en
 \bea
 \frac{T}{H_0}=\frac{A}{2\pi H_0}&=&\left| \frac{x\Omega_m}{2\pi\sqrt{\Omega_{r_c}}}
 (1+3w/2)+\frac{1}{4\pi\sqrt{\Omega_{r_c}}}
 \left[\frac{X}{H_0^2}-3\Omega_m x(1+w)
 \frac{1}{1+\frac{\epsilon}{\sqrt{X/H_0^2}}\sqrt{\Omega_{r_c}}}
 \right] \right|, \nonumber \\
 &=&\left| \frac{x\Omega_m}{2\pi\sqrt{\Omega_{r_c}}}
 (1+3w/2)+\frac{1}{4\pi\sqrt{\Omega_{r_c}}}
 \left[\frac{4\pi^2 T'^2}{H_0^2}-3\Omega_m x(1+w)
 \frac{1}{1+\frac{\epsilon}{\sqrt{T'/H_0}}\sqrt{\Omega_{r_c}}}
 \right] \right|
  \ena
 where $H_0$ is the present value of the Hubble parameter, $w$ is
 the parameter of state equation of the matter confined to the brane,
 $x\triangleq \rho/\rho_0$, and
 $\Omega_m, ~\Omega_{r_c}$ are defined as
 \bea
 \Omega_m=\frac{\rho_0}{3\mu^2H_0^2},\\
 \Omega_{r_c}=\frac{1}{r_c^2H_0^2}.~~
 \ena
 Here $\rho_0$ denotes the present density. First, we study various
 limits of the two types of the temperatures. When $r_c \to \infty,~
 \Omega_{r_c}\rightarrow 0$, we expect that DGP brane theory
 reduces
 to standard 4-dimensional cosmology. It is right in the case for
 geometric temperature of the brane,
 \be
 \lim_{r_c\to \infty}T'=\frac{\sqrt{\rho}}{2\pi\sqrt{3}\mu}.
 \en
 But for Unruh temperature, it is a different situation. Though all
 dynamical effect of the 5th dimension vanishes when $r_c\to \infty$,
 the Unruh temperature does not vanish. One can check
 \be
 \lim_{r_c\to \infty}T=\frac{1}{2}\left|5+3w\right| \frac{\sqrt{\rho}}{2\pi\sqrt{3}\mu}.
 \en
 This result implies that even when the gravitational effect of the extra
 dimension vanishes, the quantum effect is saved. The other important
 limit is the limit when the matter on the brane is infinitely
 diluted, i.e., $\rho\to 0$,
 \bea
 \label{limT'}
 \lim_{\rho\to
 0}T'=\frac{\sqrt{2}}{2\pi}\frac{\sqrt{1+\epsilon}} {r_c}, \\
 \lim_{\rho\to 0}T=\frac{1}{2\pi}\frac{1+\epsilon}{r_c}.~~~~
 \label{limT}
 \ena
 One may conclude
 \be
 \lim_{\rho\to
 0}T'=\lim_{\rho\to
 0}T,
 \en
 for either $\epsilon=1$ or $\epsilon=-1$. However
 \be
 \lim_{\rho\to
 0}\frac{T'}{T}=1,
 \en
 for $\epsilon=1$, and
  \be
 \lim_{\rho\to
 0}\frac{T'}{T}=\frac{1}{|2+3w|},
 \en
 for $\epsilon=-1$, since the ``speeds" are different when $T$ and
 $T'$ go to zero.

 An important case for cosmology is a universe filled with dust
 matter, i.e., $w=0$. In this case, (\ref{gt}) and (\ref{ut})
 become,
 \be
 T'=Z+\epsilon Y,
 \label{t'mn}
 \en
 and
 \be
 T=\left|Y+\epsilon Z+\frac{3}{2} \frac{Z^2-Y^2}{2Y+\epsilon Z}\right|,
 \label{tmn}
 \en
 where we set
 \be
 Y=\frac{1}{2\pi} H_0\sqrt{\Omega_{r_c}}~,
 \en
 and
 \be
 Z=\frac{1}{2\pi} H_0\sqrt{\Omega_{r_c}+x\Omega_m}~.
 \en
  For a reasonable cosmological model, we have $Y>0,~Z>0$ and $Z>Y$.
 The calculation to take absolute value is difficult to deal with.
 Hence we consider different cases in which we can determine the
 sign in the absolute calculation.

 First, in the branch $\epsilon=1$, $T$ becomes
 \be
  T=Y+Z+\frac{3}{2} \frac{Z^2-Y^2}{2Y+ Z},
  \en
  because every term is larger than zero. Then
  \be
  T-T'=\frac{3}{2} \frac{Z^2-Y^2}{2Y+ Z}>0.
  \en
  Therefore, the Unruh radiation perceived by a brane is warmer than
  the brane measured by the geometric temperature.

  Second, we consider the branch $\epsilon=-1$. The Unruh
  temperature in this case is much more complicated. First, at the
  point $2Y=Z$, or equally $3\Omega_{r_c}=x\Omega_m$, there is a
  singularity $T\to \infty$ in (\ref{tmn}), which means the
  particle numbers are equal at every energy level. This is valid
  only for a system endowed with finite energy levels, which
  indicates our theory must be cut off at some energy level. On the
  right hand of this singularity $2Y-Z>0$,
  $T$ can be decomposed in the form
  \be
  T=\left|\frac{(Y-Z)(Y-5Z)}{2(2Y-Z)}\right|,
  \en
  where $Y-Z<0$,  $Y-5Z<0$ and $2Y-Z>0$, hence $\frac{(Y-Z)(Y-5Z)}{2(2Y-Z)}>0$. So we remove
  the absolute sign directly and compare $T$ and $T'$ by
  \be
  T-T'=\frac{(Y-Z)(5Y-7Z)}{2(2Y-Z)}>0,
  \en
  since $Y-Z<0$, $5Y-7Z<0$ and $2Y-Z>0$. On the left hand of this
  singularity $2Y-Z<0$, similar to the case $2Y-Z>0$, we can prove
  $\frac{(Y-Z)(Y-5Z)}{2(2Y-Z)}<0$. Hence we remove the absolute sign
  by inserting a minus sign and, $T-T'$ becomes
  \be
  T-T'=-\frac{3}{2}\frac{Z^2-Y^2}{2Y-Z}>0.
  \en
  Thus we prove that the Unruh radiation perceived  by a dust
 dominated brane is warmer than the brane measured by the geometric
 temperature all the time.

 The other point
 deserving to be noted is that $T'$ will be always equal to $T$ if the parameter
 of state equation of the matter is confined to brane $w=-1$. In this
 case,
 \be
 T=T'=\frac{1}{2\pi}\left(\epsilon\frac{1}{r_c}+\sqrt{\frac{1}{r_c^2}+
 \frac{\rho}{3\mu^2}}\right).
 \en

  \begin{figure*}
\centering
 \includegraphics[totalheight=2.5in, angle=0]{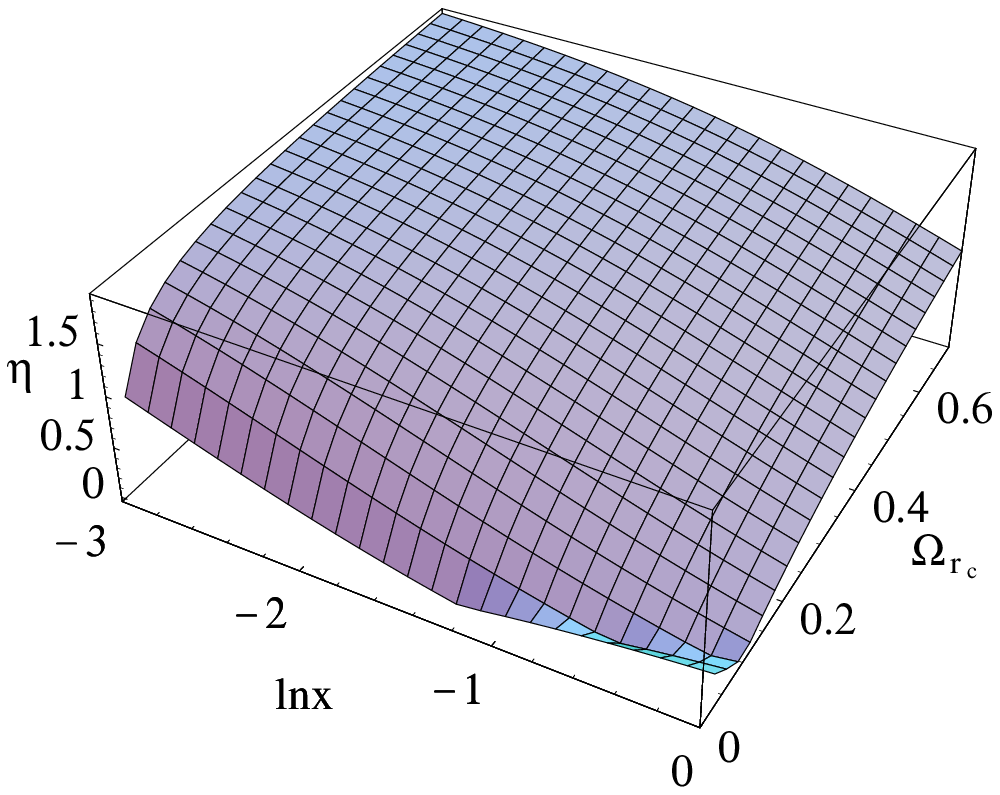}
\includegraphics[totalheight=2.5in, angle=0]{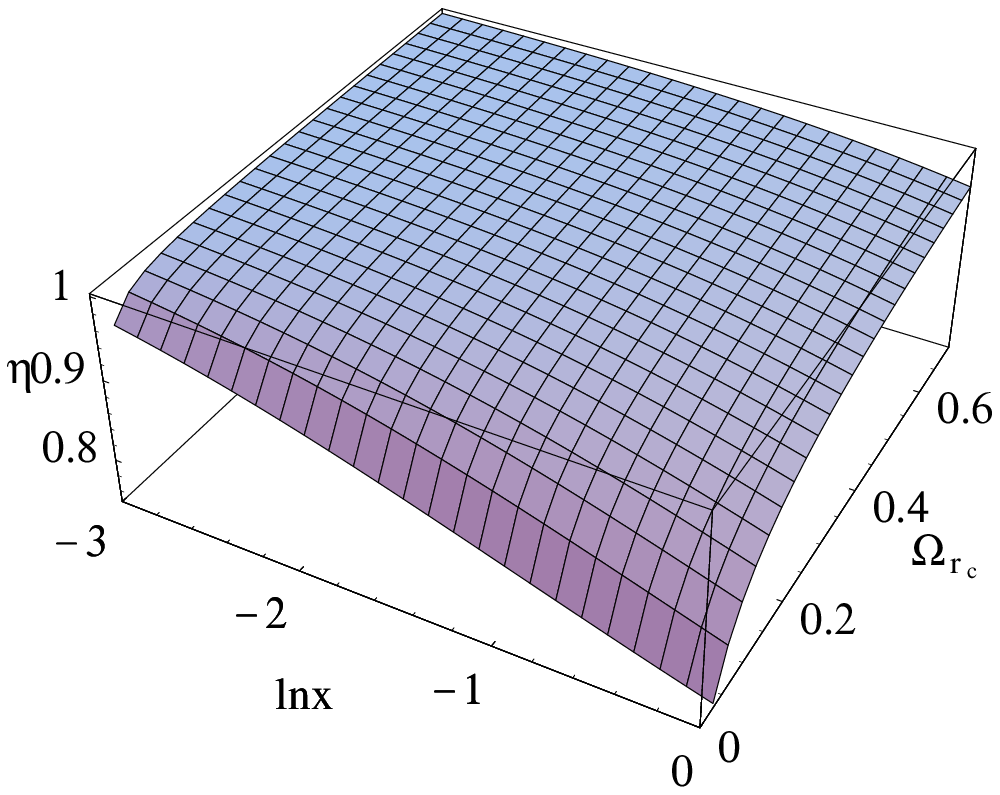}
\caption{$\eta\triangleq T'/T$ as a function of $x$ and
$\Omega_{r_c}$. In this figure we consider quintessence-like matter
dominated universe, in which $w=-1/2$ and we set $\Omega_m=0.3$.
{\bf{(a)}} The branch $\epsilon=-1$. {\bf{(b)}} The branch
$\epsilon=+1$.}
 \label{tt'}
 \end{figure*}

 \begin{figure*}
\centering
\includegraphics[totalheight=2in, angle=0]{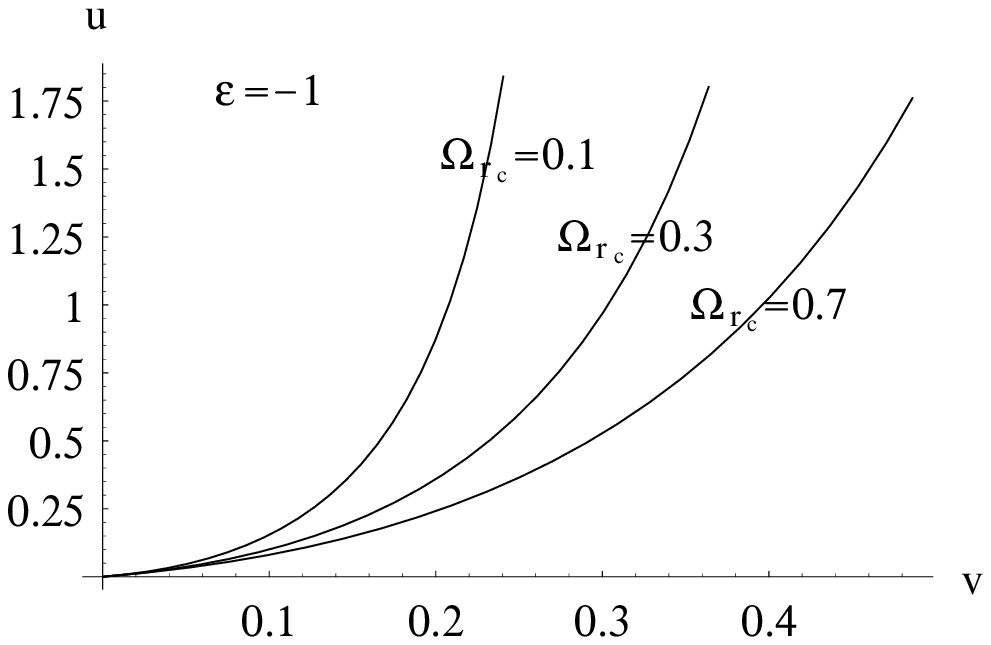}
\includegraphics[totalheight=2in, angle=0]{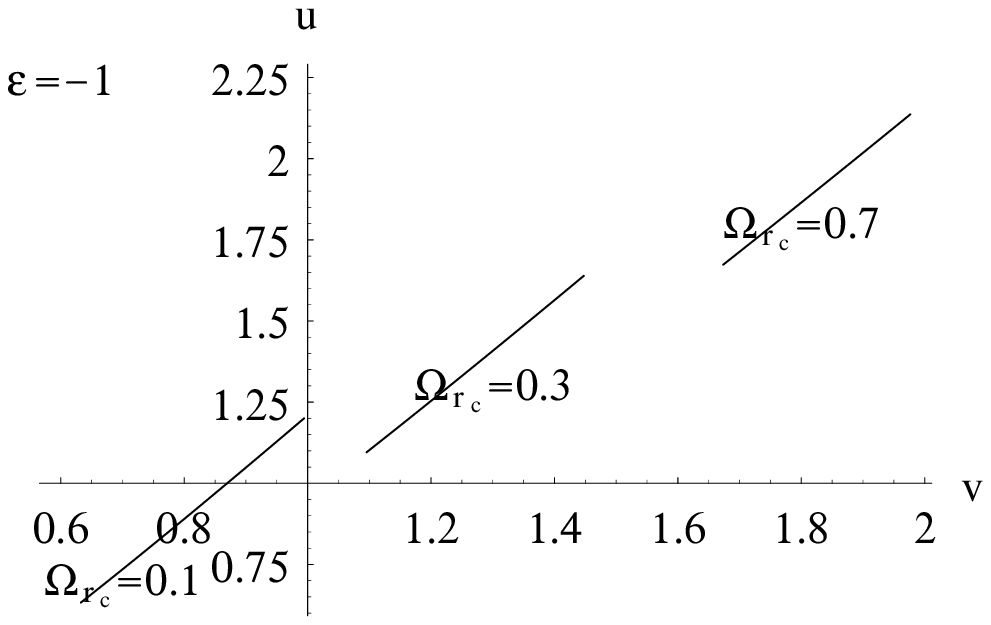}
\caption{$u\triangleq 2\pi T/H_0$ as a function of $v\triangleq 2\pi
T'/H_0$. In this figure we also consider quintessence-like, in which
$w=-1/2$ and we set $\Omega_m=0.3$. {\bf{(a)}} The branch
$\epsilon=1$. {\bf{(b)}} The branch $\epsilon=-1$.}
 \label{2dim}
 \end{figure*}

 For general case the expressions of $T$ and $T'$ are rather complicated.
 Hence we plot two figures to give their visual profiles. Fig. 1
 illustrates $T'/T$ as a function of $x$ and
$\Omega_{r_c}$. As we have pointed out, when $\rho\to 0$, $T'/T\to
 1$ for the branch $\epsilon=1$; while when $\rho\to 0$, $T'/T\to
 1/|2+3w|=1/2$ for the branch $\epsilon=1$. Fig. 2 directly displays
 $T$ as a function of $T'$, in which the interval of the argument $x$ is
 $x\in (0,3)$. Recalling $x=\rho/\rho_0$, we see that we consider the temperatures in some low redshift region and in the
 future in figure 2.
  It is clear that $T\to 0$ when $T' \to 0
 $, as shown by (\ref{limT'}) and (\ref{limT}) for the branch $\epsilon=1$.
 Our numerical
 result also shows an interesting property of the branch $\epsilon=-1$:  $T$ is
 almost a linear function of $T'$.

 \section{Unruh radiation accelerates the universe}

   With the detailed studies in the last sections, we see that the Unruh temperature does not vanish
  even at the limit $r_c\to \infty$, which is the condition for the vanishing of  the gravitational
  effect of the 5th dimension. Furthermore, the Unruh temperature is
 always higher than geometric temperature for a dust dominated brane in the
 whole history of the universe for both of the two branches and for all
 three types of spatial curvature. So an influx from bulk to brane
 will appear if brane interacts with bulk based on the most sound
 principle of thermodynamics.

  Although this difference in temperature
 indicates an energy flux, it offers no hint of the form of the
 interaction term. Under this situation, we explore the microscopic mechanism
  based on statistical mechanics and particle physics to derive the
  interaction term between Unruh radiation and dark matter confined
  to the brane.

  In section II, we have considered a detector coupling to a scalar field
  $
 L_{\phi}=\frac{1}{2}\partial_{A}\phi\partial^A\phi-\frac{1}{2}m^2\phi^2
 $, which satisfies
  5 dimensional Klein-Gordon equation. At the massless limit, it
  obeys
  \be
  \square^{(5)} \phi=0.
  \label{5kg}
  \en
    By using Green function method, we derive the temperature of
  the Unruh radiation,
 \be
 T=\frac{A}{2\pi},
 \label{tem1}
 \en
 where $A$ is given by (\ref{acce}).

 To investigate the microscopic mechanism of interaction between brane and bulk,
 one should decompose the 5-dimensional modes, which satisfies (\ref{5kg}),
  into modes along the brane and modes transverse   to the brane.
  Following \cite{yling}, we call them type I modes and type II
  modes, respectively. In a pseudo-Euclidean coordinates system,  type I mode reads,
  $\phi^{(I)}=e^{-ik_\mu x^\mu}$,
  where $k_\mu$ denotes the momentum of the mode in different directions,
  and  $\mu=0,1,2,3$. Type II modes reads,
 $ \phi^{(II)}=e^{-ik_\mu x^\mu-ik_y y}$,
    where $y$ is the coordinate of the extra dimension, and we require $k_y\neq 0$.
  An arbitrary 5 dimensional mode must be either type I or type II.
  It was shown in \cite{yling}
  that
  Type II modes of the bulk  fluctuations do not interact with the brane,
   for which the brane is effectively a mirror.
  Hence only type I mode can
   exchange energy
   with matter confined to the brane. We show a schematic plan of
   type I and type II modes in Fig. 3.

   \begin{figure}
\centering
\includegraphics[totalheight=2.7in, angle=0]{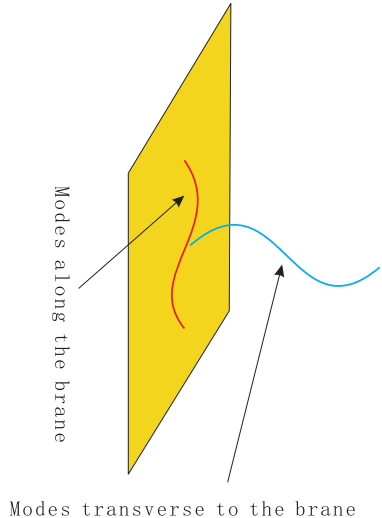}
\caption{A sketch of type I and type II modes. Only type I mode can
 exchange energy with particles confined to the brane, while the brane
 behaves as a perfect mirror for type II mode (for a color version on line).}
 \label{fig2dim}
 \end{figure}

 By a standard process of canonical quantization, type I modes
 become photons on the brane. Here photons mean quanta without
 mass, which do not necessarily satisfy Maxwell equations. In fact we
 do not even need to study  its commutativity in detail. Whether they obey
 Fermi-Dirac statistics or Bose-Einstein statistics the number density
 of the photon gas $n_\gamma$ is proportional to the cubic of its temperature,
 \be
 \label{ngamma}
 n_\gamma \propto T_\gamma^3,
 \en
 where $T_\gamma$ is just the temperature of the Unruh radiation
 (\ref{tem}).
 The particles confined to the brane are immerged in the thermal bath
 of these photons. Therefore, based on statistical mechanics the reaction
 rate $\Gamma$ between dark matter particles and the type I
 photons is proportional to the number densities of photons and dark matter
 particles, the relative velocity of the photons and matter
 particles,
  and the scattering cross section $\Gamma \propto
  n_{\rm dm}n_{\gamma}\sigma v_{\rm pm}$. Here,
    $n_{\rm dm}$ denotes the number density of
 dark matter particle, $v_{\rm pm}$ stands for the relative velocity, which is a constant, and $\sigma$ represents the scattering cross
 section. In a low energy region, the internal
 freedoms can not be excited. This paper concentrates on the late
 time universe, hence the cross section is effectively constant.
 Therefore, the reaction rate can be written as an equation by inserting a constant $b$,
 \be
 \Gamma=b\rho A^3,
 \label{gamma}
 \en
 where we have used (\ref{ngamma}), (\ref{tem}) and $n_{dm}\propto \rho$.
  Here, we assume there is only pressureless dark
 matter on the brane, which interacts with the Unruh photons.
 Therefore, the continuity equation of the brane becomes,
 \be
 \dot{\rho}+3H\rho=\Gamma.
 \label{conti}
 \en

 As we explained before, the DGP braneworld model also suffers from
 fine-tune problem, which says why the three RHS terms of (\ref{fried}) are
 at the same scale.  To overcome this hurdle, we consider a
 limiting DGP model, that is, $r_c\gg H_0^{-1}$. In such a model,
 the fine-tuned problem is evaded, and at the same time the
  gravitational effect of the 5th dimension is no longer responsible for
  the present acceleration. We have seen that the
  Unruh effect does not vanish even when $r_c\to \infty$. We will
  prove that under the situation $r_c\gg H_0^{-1}$ the interaction between dark
  matter on the brane and Unruh radiation can drive the observed acceleration of
  the universe.

  In a limiting DGP model
   , substituting the Friedmann equation (\ref{fried}) into (\ref{conti}), we derive
  \be
  \frac{\ddot{a}}{{a}}=-\frac{\rho}{6\mu^2}\left[1-b A^3\left(\frac{\rho}{3\mu^2}-\frac{k}{a^2}\right)^{-1/2}\right].
  \label{dda}
  \en
  The acceleration $A$ of the brane in the bulk  does not directly
  depend on energy density of the matter on the brane $\rho$, but through
  the effective density $\rho_e$ and and effective pressure $p_e$ in
  (\ref{acce}), which are presented in (\ref{effrho}) and
  (\ref{effp}), respectively.

  Associating (\ref{acce}), (\ref{effrho}),
  (\ref{effp}),  and (\ref{dda}) we
  derive
  \be
   A^{-2}=\frac{\kappa^2}{6}b\rho
   \left(\frac{\rho}{3\mu^2}-\frac{k}{a^2}\right)^{-1/2}.
   \label{A}
   \en
    The formula (\ref{A}) is general for
  all the three cases of curvatures. In the case of a spatially flat brane, it
  takes a simple form,
  \be
   A^{-2}=C\sqrt{\rho},
   \label{inver}
   \en
  where $C=\kappa^2\mu b$. Interestingly, we see that the acceleration of the brane $A$
  inversely correlates to the energy density of the brane, which is
  completely different from the first sight at equation
  (\ref{acce}). Therefore, we expect the effect of the bulk Unruh radiation
  is negligible in the early time for a dust dominated limiting DGP
  braneworld. Only in some low energy region the bulk Unruh
  radiation becomes important.

  Before presenting the exact cosmic solution, we study some
  qualitative side of the Firedmann equation (\ref{fried}) and continuity
  equation (\ref{conti}) to see whether our model is stable. We consider a spatially flat universe, in
  which the stagnation point dwells at $\delta H=0$, or equivalently
  $\delta \rho=0$. From the continuity equation (\ref{conti}), we
  derive the energy density $\rho_s$ at the stagnation point,
  \be \rho_s^{5/4}={b\mu C^{-3/2}}/{\sqrt{3}}. \en
  To investigate the  stability of the cosmic fluid in the
  neighbourhood of the stagnation point, we impose a perturbation
  to the continuity equation,
  \be
  (\delta \rho)^{.}=\delta \rho \left(\frac{1}{4}b \rho^{-3/4}C^{-3/2}
  -\frac{3\sqrt{3}}{2\mu}\rho^{1/2}\right).
  \en
   At the stagnation point, $\rho=\rho_s$, hence
   \bea
    (\delta \rho)^{.}|_{\rho=\rho_s}
  =-\frac{5\sqrt{3}}{4\mu}\rho_s^{1/2}
  \delta \rho,
  \ena
  which means it is a stable point.
  Now we consider the deceleration parameter, which is one the most
  significant parameters from the viewpoint of  observations. Here
  the deceleration parameter reads,
  \bea
  \label{dece}
   q
   =\frac{1}{2}\left(1-\frac{b
   A^{3}}{\sqrt{\frac{\rho}{3\mu^2}-\frac{k}{a^2}}}\right)\left(1-\frac{3\mu^2}{\rho}\frac{k}{a^2}\right)^{-1},
   \ena
   where $A$ is given by (\ref{A}).
   In the case of a  spatially flat universe, it degenerates to
  \bea
  \label{deceflat}
   q
   =\frac{1}{2}\left(1-\sqrt{3}\mu b C^{-3/2}\rho^{-5/4}\right).
   \ena
  This equation clearly shows that in the early universe $q\to 1/2$,
  hence the universe behaves as dust dominated one, and with the decreasing of
  energy density the deceleration parameter becomes smaller.
  Finally at the stagnation point the deceleration parameter ceases
  at
  $q=\frac{1}{2}\left(1-\sqrt{3}\mu b
  C^{-3/2}\rho_s^{-5/4}\right)=-1$,
  which implies that the universe enters a de Sitter phase.

  Though, for an arbitrary spatial curvature the analytical
  solution does not exist, we find an exact solution for a spatially
  flat universe driven by bulk Unruh radiation. For a spatially flat
  universe,
  associating Friedmann equation (\ref{fried}) with the continuity equation (\ref{conti})
 we obtain,
 \be
 \rho=\left(c_1a^{-15/4}+\frac{L}{3}\right)^{4/{5}},
 \label{exactrho}
 \en
 where $c_1$ is an integration constant, and $L$ is defined by
 $
 L=\sqrt{3}\mu bC^{-3/2}
 $.
 When $c_1>0$, from this exact solution, the universe behaves as
 dust dominated one in a high energy region ($a$ small enough), and
 becomes de Sitter universe in a low energy region ($a$ large
 enough), which is exactly the same as we concluded before from behaviors of the
 deceleration parameter in the history of the universe. When $c_1<0$,
 there is a bounce in the early universe, and the universe also
 enters a de Sitter phase in a low energy region. Since our
 investigations for the acceleration driven by Unruh radiation
 concentrate on the low energy region, it is only a toy model in the
 early universe.

  With the exact solution (\ref{exactrho}) and
 (\ref{fried}), we obtain an analytical expression of cosmic time $t$
 as function of the scale factor $a$ by using a hypergeometric function,
 \bea
 \frac{t}{\sqrt{3}\mu}+c_2=
 \frac{2}{3} \frac{a^{3/2}}{c_1^{2/{5}}}F(\frac{2}{5},\frac{2}{5},\frac{7}{5},-a^{15/4}\frac{L}{3c_1})
 ,
 \label{exact}
 \ena
 where $c_2$ is an integration constant, and $F$ denotes Gauss
 hypergeometric function. The requirement
 \be
 \lim_{a\to 0}t=0,
 \en
 yields $c_2=0$.

  To visualize the physical meaning of this solution, we demand the various
 limits
 of it. In the high energy limit (small $a$), expanding (\ref{exact})
 in series around $a=0$,
 \be
  \frac{t}{\sqrt{3}
  \mu}=\frac{2}{3c_1^{2/5}}a^{3/2}-\frac{8L}{315c_1^{7/5}}a^{21/4}+
  {\cal{O}}(a^{17/2}),
  \en
  where if we only keep the dominated term we just obtain $a\sim
  t^{2/3}$, as we expected. In a low energy region when $a$ is large
  enough, the series becomes,
  \bea
  \nonumber
   \frac{t}{\sqrt{3}\mu} =\frac{4}{5}
   \frac{1}{3^{3/5}L^{2/5}}\left[-\gamma+\ln(\frac{L}{3c_1})+\frac{15}{4}\ln
   a-\psi(\frac{2}{5})\right] \\
   +\frac{3^{2/5}}{25}\frac{8c_1a^{-15/4}}{L^{7/5}}+{\cal
   O}(a^{-29/4}),~~~~~~~~~~~~~~~~~~~~~~
   \ena
  where $\gamma$ is the Euler constant, and $\psi$ is the digamma
  function. If we only take the leading term, we obtain
  $a= Me^{Nt}$, where
  $M=\exp\{\frac{4}{15}[\gamma-\ln(L/3c)+\psi(2/5)]\}$,
  $N=L^{2/5}3^{3/5}(3\sqrt{3}\mu)^{-1}$, which is just a de Sitter space.

 Thus, we complete a  cosmic
 solution of the limiting DGP braneworld model,
 where the universe is self-accelerated through the bulk Unruh
 radiation perceived by the brane. From the viewpoint of
 observations
 our model belongs to the class of
 unified dark energy model, that is, there is only one component (usually dust) in
 the universe, but due to different reasons it evolves in
 a non-standard way. Therefore, it can drive the present acceleration of
 the universe. Such models also have been phenomenologically investigated in, for example, \cite{chap}.

 Finally, we would like present some preliminary discussions to confront the
 observations. First, the parameter $b$  is critical to our model, which
 encloses all the undetermined information of interaction between
 dark matter and Unruh radiation. Here we  present a
 preliminary estimation of its value from the deceleration parameter
 (\ref{deceflat}) in a spatially flat universe. From various observations the present value of
 deceleration parameter $q\sim -0.5$, therefore,
 \be
  \frac{\sqrt{2}}{  \sqrt{bH_0^2}}(r_cH_0)^{-3/2}\left(\frac{\rho}{3\mu^2
 H_0^2}\right)^{-5/4}\sim 1,
 \en
 where $\frac{\rho}{3\mu^2
 H_0^2}=1$ in our model. We see that a larger cross radius $r_c$ needs a smaller
 coupling constant $b$. And we have set $r_cH_0\gg 1$ in previous
 constructions. Hence the dimensionless coupling constant $bH_0^2$
 is a tiny number, which eludes our laboratory experiments even if the
 dark matter particles were found. Alternatively, in future work we expect
 astronomical observations to fit our model and hence to determine the
 value of $b$, which is also helpful to constrain
 the cross radius $r_c$ in DGP model.

    An exotic matter with negative pressure, call dark energy, is frequently introduced to
  explain the cosmic acceleration in frame of general relativity.
    To explain observed accelerated expansion, we calculate the equation of state $w$ of the effective
  ``dark energy" caused by the induced Ricci term and energy influx
  from the bulk Unruh radiation
  by comparing the modified Friedmann equation in
  the brane world scenario and the standard Friedmann equation in general
  relativity, since almost all observed properties of dark energy are
  ``derived" in general relativity.
  The Friedmann equation in the
four dimensional
  general relativity can be written as
 \be
 H^2+\frac{k}{a^2}=\frac{1}{3\mu^2} (\rho+\rho_{de}),
 \label{genericF}
 \en
 where the first term of RHS of the above equation represents the dust matter and the second
 term stands for the dark energy.
 Generally speaking the Bianchi identity requires,
  \be
 \frac{d\rho_{de}}{dt}+3H(\rho_{de}+p_{de})=0,
 \label{em}
 \en
 we can then express the equation of state for the dark
 energy as
   \be
  w_{de}=\frac{p_{de}}{\rho_{de}}=-1-\frac{1}{3}\frac{d \ln \rho_{de}}{dlna}.
  \label{wde}
   \en
   Comparing (\ref{genericF}) and (\ref{fried}), we derive
   \be
   \rho_{de}=\frac{2}{r_c^2}+\frac{2\epsilon}{r_c}
 \left(\frac{\rho}{3\mu^2}+\frac{1}{r_c^2}\right)^{1/2}.
 \label{rhode}
  \en
  Note that in our model there is no exotic matter confined to the
   brane. $\rho_{de}$  is in fact geometric and interaction effect.
   We call $\rho_{de}$ equivalent or  virtual density of dark
   energy.
 Various evidences, which are independent to cosmological models, implies the existence
  of dark matter with present density about $\Omega_{m}= 0.2\sim 0.4$ \cite{darkmatter}. Maybe more or less, but the density
  of dark matter does reach the density to flat the space. In our
  model, the geometric contribution of $r_c$ can be very small. So
  generally speaking, we need a curvature term. Substituting
  (\ref{rhode}) into (\ref{wde}), and recalling that the matter confined to the brane is pressureless,  we obtain,
  \be
  w_{de}=-1+\frac{\epsilon}{3}\frac{\dot{\rho}}{3\mu^2H}\left(\frac{\rho}{3\mu^2}+\frac{1}{r_c^2}\right)^{-1/2}
  \left[\frac{2\epsilon}{r_c}\left(\frac{\rho}{3\mu^2}+\frac{1}{r_c^2}\right)+\frac{2}{r_c^2}\right]^{-1},
    \en
   where $\dot{\rho}$ is given by (\ref{conti}), and $\Gamma$ can be
   calculated by (\ref{gamma}). Far a large $r_c$, after complicated
   but straightforward algebraic calculating, we deduce,
   \be
   w_{de}=-\frac{3}{2}+Q\Omega_m^{-3/2}(1+z)^{-9/2}\left(\Omega_m(1+z)^3+\Omega_k(1+z)^2\right)^{1/4},
      \en
   where $Q$ is defined as,
   \be
   Q=\frac{18^{3/2}}{6}(r_cH_0)^{-3/2}(bH_0^2)^{-1/2},
   \en
      \be
   \Omega_k=\frac{-k}{H_0^2a^2},
   \en
  and $z$ is the redshift.
   It is clear that when $Q\Omega_m^{-3/2}(\Omega_m+\Omega_k)^{1/4}>0.5$, the virtual dark energy behaves as
   quintessence; while when $Q\Omega_m^{-3/2}(\Omega_m+\Omega_k)^{1/4}<0.5$, it behaves as
   phantom. We stress for the second time that the dark energy in this model is
   only some virtual, not actual stuff. The most sensible quantity
   in observation is the deceleration parameter $q$, which is given
   in (\ref{dece}). Substituting (\ref{A}) into (\ref{dece}), we
   reach,
   \be
   q=\frac{1}{2}-3^{-1/2}Q\Omega_m^{-3/2}(1+z)^{-9/2}\left(\Omega_m(1+z)^3+\Omega_k(1+z)^2\right)^{1/4},
   \en
  where we also supposed a large $r_c$. Note that for both $q$ and
  $w_{de}$, $\epsilon$ vanishes at the large $r_c$ limit. This
  reasonable, since the two branches have no difference at such a
  limi.t

\begin{figure}
\centering
\includegraphics[totalheight=2.7in, angle=0]{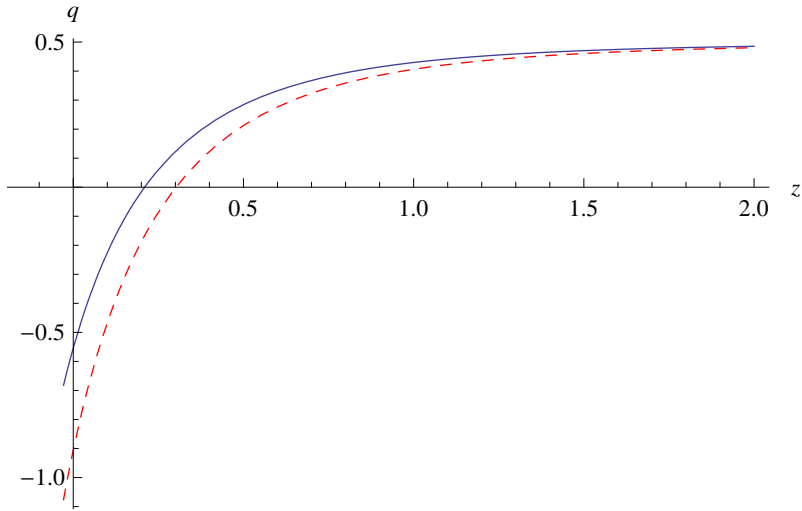}
\caption{The evolution of deceleration parameter with respect to
$z$. For both of the two curves, $\Omega_m=0.3$, $\Omega_k=0.7$.
$Q=0.4$ on the solid curve, and $Q=0.3$ on the dashed curve.}
 \label{figdece}
 \end{figure}

  As numerical examples, we plot a figure for $q$ for two sets of
  parameters in fig \ref{figdece}. This figure illuminates that the
  universe is accelerating and $q$ goes to $0.5$ quickly at
  high redshift region, ie, the universe becomes a dust one.

 \section{conclusion}
 Our scenario is that the brane is accelerating in the bulk, such
 that it is bathed in the Unruh radiation in the bulk. The temperature of the
 Unruh radiation is higher than the brane, which yields energy flux
 from the Unruh radiation to the brane matter, which accelerates the
 brane universe.

   In this article we study the
  cosmology case of a DGP braneworld model, that is, an FRW brane imbedded in the bulk spacetime.  We
  first investigate the Unruh temperature that the brane perceives and its ``own" temperature.
  Generally speaking the
 brane accelerates in the bulk. Hence the particles confined to the brane and inertial on the brane
  should also perceive Unruh radiation in the bulk. In this paper we show that for a DGP brane
 in a Minkowski bulk it is just the case.
  We investigate the case of the brane, as a
 particle detector, coupled to a massless scalar field in the bulk.
  As for a point detector, the temperature of the radiation
  perceived is proportional to the acceleration of the brane.

   As for the brane's own temperature, we should seek a
  characteristic temperature of the brane. The temperature can reveal the inherent property of gravity.
   The Friedmann equation can be
  reproduced by applying the first law of thermal dynamics to the apparent horizon.
  Thus we take the temperature of the apparent horizon as the characteristic temperature
  of the brane, which we called geometric temperature.

  We find that generally speaking the Unruh temperature the brane perceived and geometric
  temperature
  of the brane are not equal. We   compare these two temperatures in various cases.
  We find that for a dust dominated brane,
 the temperature of the Unruh radiation perceived by a brane is always higher than the geometric
 temperature of the brane, either in the branch $\epsilon=1$ or $\epsilon=-1$, no matter what the value of the cross
 radius $r_c$ and the spatial curvature of the brane. So, generally
 speaking the brane and the Unruh radiation it perceived can not
 reach to thermal equilibrium.

  As we pointed out
 before, the validity of Unruh radiation has been confirmed in detail \cite{review},\cite{lin}.
 Therefore if energy exchange is allowed between bulk and matter brane, an
 energy flux between
 the bulk radiation and the matter confined to the brane will
 come forth, which may accelerate our universe.
  In all cases $\rho>0$, the Unruh temperature of
 the bulk radiation $T$ is higher than than the
 geometric temperature $T'$ for a dust dominated brane, which means an energy influx
 to the brane can appear. So we study this possibility in section V.

 The DGP braneworld model seems to be a hopeful candidate to explain the
 cosmological acceleration. But, as we pointed out before, it
 also suffers from fine-tune problem. To evade this problem, we
 consider the limiting DGP model, that is, $r_c>H_0^{-1}$, which
 means the gravitational effect of the 5th dimension is negligible.
 Under this condition, the cosmological acceleration does not happen in original DGP model
 under this situation if there is only dust on the brane.
  By considering the possible energy influx from the bulk Unruh radiation to the brane induced by the
  temperature grads, we find the universe can
 accelerate through interacting the bulk Unruh radiation in a limiting DGP model,
  even the
 brane is dust dominated. Differently from previous works on the brane-bulk interaction, we find the
 interaction form through careful
 studies on the microscopic mechanism of interaction between brane and bulk.
 It is shown that the interaction term can be settled up to a
 constant factor $b$. Based on these constructions we find the acceleration of the brane
 $A$ is inversely correlated with the energy density of the brane for
 a dust dominated limiting DGP brane.  Finally we derive an
 exact solution for a spatially flat model. This solution shows
 clearly that the universe behaves as a dust dominated one at early
 time and enters a de Sitter phase at late time, which is
 consistent with observations. We also show the de Sitter phase is
 stable.

 In this paper only massless Unruh mode is considered. Although it is the
 most important mode in low energy region, the massive mode also
 deserves to study further for a full Unruh effect in a high energy region.
 Also, in the enough high energy region, the universe is radiation
 dominated, and at the same time the scattering cross section between dark matter particles
 and Unruh photons  becomes temperature-dependent, since the internal freedoms can
 be excited. Under this
 situation the Unruh effect may be important again, which deserves to
 investigate further.

 At the observation side, we should not only take a special set of
 parameters to show the property of this model, but constrain the
 parameters, especially $b$ by various observations in the future.


 {\bf Acknowledgments.}
 HS Zhang thanks Prof. W. Unruh and Prof. D. Jennings  for helpful discussions
 . H. Noh was supported by grant No. C00022 from the Korea Research
 Foundation. ZH Zhu was supported by the National Science Foundation of China under
 the Distinguished Young Scholar Grant 10825313, the Key Project Grant 10533010,
and by the Ministry of Science and Technology national
 basic science Program (Project 973) under grant No. 2007CB815401. HW Yu was supported by the National
 Natural Science Foundation of China  under Grants No. 10575035 and
  No. 10775050, the SRFDP under
 Grant No. 20070542002,  and the
 Program for the key discipline in Hunan Province.


\begin{thebibliography}{99}

 \bibitem{acce}
  A. G. Riess et al.,
  Astron. J. 116, 1009 (1998);
  S. Perlmutter et al.,
  Astrophys. J. 517, 565 (1999).



\bibitem{review}
 E.~J.~Copeland, M.~Sami and S.~Tsujikawa,
  Int.\ J.\ Mod.\ Phys.\  D {\bf 15}, 1753 (2006)
  [arXiv:hep-th/0603057].

  \bibitem{rev}
  R.~Maartens,
  Living Rev.\ Rel.\  {\bf 7}, 7 (2004)
  [arXiv:gr-qc/0312059].




 \bibitem{dgpmodel}
 G. Dvali, G. Gabadadze, M. Porrati, Phys. Lett. B485 (2000) 208
 ,hep-th/0005016; G. Dvali and G. Gabadadze, \prd
 {\bf 63},
 065007 (2001); A. Lue, astro-ph/0510068.

\bibitem{dgpcosmology}
C. Deffayet, Phys. Lett. B 502, 199 (2001); C. Deffayet, G.Dvali,
and G. Gabadadze, Phys. Rev. D 65, 044023 (2002); C.Deffayet, S. J.
Landau, J. Raux, M. Zaldarriaga, and P. Astier, Phys. Rev. D 66,
024019 (2002);


 \bibitem{interbrane}
  E. Kiritsis, G. Kofinas, N. Tetradis, T.N. Tomaras and V. Zarikas,
  JHEP 0302 (2003) 035;
    C. van de Bruck, M. Dorca, 10 C. J. Martins
 and M. Parry, Phys. Lett. B 495 (2000) 183 ;
 N. Tetradis, Phys.Lett. B569 (2003) 1;
  F.~K.~Diakonos, E.~N.~Saridakis and N.~Tetradis,
  %
  Phys.\ Lett.\ B {\bf 605}, 1 (2005);
    E. Kiritsis, JCAP 0510, 014 (2005)
; K. I. Umezu, K. Ichiki, T. Kajino, G. J.
 Mathews, R. Nakamura and M. Yahiro, \prd 73, 063527 (2006); G.
 Kofinas, G. Panotopoulos and T.N. Tomaras;
 R.~G.~Cai, Y.~g.~Gong and B.~Wang,
  %
  JCAP {\bf 0603}, 006 (2006).

 \bibitem{jenn}
 D. Jennings, hep-th/0508215.

\bibitem{4law}
  J.~M.~Bardeen, B.~Carter and S.~W.~Hawking,
  Commun.\ Math.\ Phys.\  {\bf 31}, 161 (1973).

  \bibitem{hawk}
  S.~W.~Hawking,
  Commun.\ Math.\ Phys.\  {\bf 43}, 199 (1975)
  [Erratum-ibid.\  {\bf 46}, 206 (1976)].


\bibitem{unruh}
 W.G. Unruh, Phys. Rev. D14, 870 (1976); W.G. Unruh,
 \prd {\bf 34}, 1222 (1986);
 W. G. Unruh, Phys. Rev. D 46, 3271 (1992).

 \bibitem{bril}
 N.D. Birrell, P.C.W. Davies, Quantum fields in curved space (1982),
 Cambrige university press, Cambrige, UK .
 \bibitem{jaco}
 T. Jacobson, Phys. Rev. Lett. 75, 1260 (1995), gr-qc/9504004.

\bibitem{cai}
  R.~G.~Cai and S.~P.~Kim,
  JHEP {\bf 0502}, 050 (2005)
  ,hep-th/0501055.
 \bibitem{bak}
 D. Bak and S. J. Rey, Class. Quant. Grav. 17, L83 (2000)
 [arXiv:hep-th/9902173];
 S. A. Hayward, S. Mukohyama and M. C. Ashworth, Phys. Lett. A 256,
347 (1999) [arXiv:gr-qc/9810006]; S. A. Hayward, Class. Quant. Grav.
15, 3147 (1998) [arXiv:gr-qc/9710089].






















 \bibitem{math}
B. S. Kay, Commun. Math. Phys. 100, 57 (1985); R.M. Wald and B. Kay,
Phys. Rep. 207, 51 ~1991; S. De Bievre, M. Merkli, math-ph/0604023.

\bibitem{no}
 P. G. Grove, Classical Quantum Gravity 3, 801 (1986);
 D. J. Raine, D.W. Sciama, and P. G. Grove, Proc. R. Soc.
A 435, 205 (1991).
 \bibitem{1992}
 W. G. Unruh, Phys. Rev. D 46, 3271 (1992).
 \bibitem{cloud}
 S. Massar, R. Parentani, and R. Brout, Classical Quantum
 Gravity 10, 385 (1993).
 \bibitem{non}
  A. Raval, B. L. Hu, and D. Koks, Phys. Rev. D 55, 4795
(1997).
 \bibitem{lin}
 S.-Y. Lin, Phys. Rev. D 68, 104019 (2003);
 S.-Y. Lin and B. L. Hu \prd 73, 124018 (2006)

\bibitem{Audretsch94} J. Audretsch and R. M\"uller, Phys. Rev. A {\bf 50},
1755 (1994).

\bibitem{ZYL06} Z. Zhu, H. Yu and S. Lu, Phys. Rev. D {\bf73},
107501 (2006).

\bibitem{ZY07} Z. Zhu and H. Yu, Phys. Lett. B {\bf645},
459 (2007).

 \bibitem{wald}
 R. Wald, Quantum Field Theory in Curved Space-
time and Black Hole Thermodynamics (1994), University of Chicago
Press, Chicago, US.
 \bibitem{yuhan}
  H.~Yu and W.~Zhou,
  Phys.\ Rev.\  D {\bf 76}, 044023 (2007)
  [arXiv:0707.2613 [gr-qc]];
 M.~Han, S.~J.~Olson and J.~P.~Dowling,
  arXiv:0705.1350 [quant-ph].
 \bibitem{kkrs}
  N.~Khosravi, E.~Khajeh, R.~Rashidi and H.~Salehi,
  Astrophys.\ Space Sci.\  {\bf 310}, 333 (2007)
  [arXiv:0706.2767 [hep-th]].




  \bibitem{ureview}
  L.~C.~B.~Crispino, A.~Higuchi and G.~E.~A.~Matsas,
  arXiv:0710.5373 [gr-qc].

\bibitem{gbzhang}
 Rong-Gen Cai, Hongsheng Zhang, and Anzhong Wang, Commun.Theor.Phys. 44
(2005) 948.

\bibitem{stable}
  C.~Deffayet, G.~Gabadadze and A.~Iglesias,
  JCAP {\bf 0608}, 012 (2006)
  [arXiv:hep-th/0607099].

 \bibitem{lang}
 D. Langlois, Prog.Theor.Phys.Suppl.
148 (2003) 181.








 \bibitem{eff}
 K. Maeda, S. Mizuno and T. Torii,
  Phys.\ Rev.\ D {\bf 68}, 024033 (2003).




 \bibitem{yling}
 S. Alexander, Y. Ling, and L. Smolin
 Phys. Rev. D 65, 083503 (2002).




\bibitem{chap}
 A. Kamenshchik, U. Moschella and V. Pasquier, Phys. Lett.
  {\bf B511} (2001) 265;  B. A. Bassett,
  M. Kunz, D. Parkinson, C. Ungarelli,Phys.Rev. D68 (2003) 043504;
  S. M. Carroll, V. Duvvuri, M. Trodden and M. S. Turner,
  Phys. Rev. D 70 (2004) 043528;O. Mena, J. Santiago and J. Weller
  Phys.Rev.Lett. 96 (2006) 041103; Hongsheng~Zhang and Z.~H.~Zhu,
  Phys.\ Rev.\  D {\bf 73}, 043518 (2006)
  [arXiv:astro-ph/0509895].

 \bibitem{darkmatter}
 K Olive, arXiv:0901.4090.


\end{thebibliography}
\end{document}